\documentclass[a4paper]{article}
\usepackage[english]{babel}
\usepackage[T2A]{fontenc}
\usepackage[utf8]{inputenc}
\usepackage{amsmath,esint}
\usepackage{amsthm}
\usepackage{amssymb}
\usepackage{textcomp}
\usepackage{graphicx}
\usepackage{array}
\usepackage{hyperref}
\usepackage{tikz}
\usepackage{float}
\usepackage[section]{placeins}
\usepackage{authblk}

\DeclareMathOperator{\sgn}{sgn}

\newcommand{\be}{\begin{eqnarray}}
\newcommand{\ee}{\end{eqnarray}}

\newcommand{\eqlb}[2]{\begin{equation} \label{#1} #2 \end{equation}}
\newcommand{\eq}[1]{\begin{equation} #1 \end{equation}}

\newcommand{\brc}[1]{\left(#1\right)}
\newcommand{\bsq}[1]{\left[#1\right]}

\newcommand{\abs}[1]{\left|#1\right|}

\newcommand{\rme}{\textrm{e}}
\newcommand{\rmd}{\textrm{d}}

\numberwithin{equation}{section}

\hoffset= - 1.0in
\begin{document}

\title{Classical string correspondence in $2$ and $3$ spacetime dimensions}
\author[a,b,c]{G. Aminov\thanks{gleb.aminov@stonybrook.edu}}
\affil[a]{\small C.~N. Yang Institute for Theoretical Physics, State University of New York, Stony Brook, NY 11794-3840, USA}
\affil[b]{\small Simons Center for Geometry and Physics, State University of New York, Stony Brook, NY 11794-3636, USA}
\affil[c]{\small ITEP NRC KI, Moscow 117218, Russia}

\date{ }

\maketitle

{\abstract We present an example of the string correspondence: a relation between two classical string solutions - one in Minkowski spacetime  and another one in the AdS spacetime. The first solution describes a $2$d motion of the string with massive ends and was derived in \cite{BBHP'76}. The second solution is the accelerating string solution in the $\textrm{AdS}_3$ spacetime and describes a heavy quark-antiquark pair \cite{Xiao'08}.}

\section{Introduction}
A notion of duality plays a prominent role in the modern theoretical physics. As the number of theoretical models grows larger and larger, one finds the need to connect them, since they sometimes describe the same physical phenomena. Some of the most famous connections are the AdS/CFT correspondence \cite{Maldacena'98,Witten'98,GKP'98}, different string dualities \cite{FT'85,Sathiapalan'87,Buscher'87,COGP'91,Sen'94,Schwarz'95,HT'95} and the AGT correspondence \cite{AGT'10,Wyllard'09}. In this paper we want to describe an example of the correspondence, that has something to do with the first two. From the pure geometrical point of view our example looks like a relation between an open string with massive ends in $2$d Minkowski spacetime and a closed string in $3$d AdS spacetime (with some specific identifications on the boundary of $\textrm{AdS}_3$). The time and space coordinates on the string worldsheet are flipped between the two solutions and thus the relation closely resembles T-duality. On the other hand, by cutting the closed string in half on the AdS side one finds a motion of the quark-antiquark pair as proposed by the AdS/CFT correspondence. Parallels with the AdS/CFT correspondence go deeper due to the fact, that the $2$d solution exhibits some form of the $\textrm{SL}\brc{2,R}$  symmetry. This could potentially allow one to define a boundary CFT and thus leads to a manifestation of AdS/CFT called AdS/BCFT \cite{Takayanagi'11,FTT'11}. Another interesting direction would be the relation between $2$d strings and matrix models \cite{BSf'92,BSc'95,Maldacena'05}. In particular, string solutions called short strings and long strings were obtained in \cite{MO'00,MOS'01,MO'02} as classical solutions of the $\mathrm{SL}(2,R)$ WZW model \cite{WZ'71,Witten'83,Witten'84,Novikov'81,Novikov'82}. These string solutions were discussed in different contexts \cite{AF'11,AK'13,BKL'21} and they are possibly related to the classical solution \cite{BBHP'76}.

The main goal of this paper is to connect two different classical solutions \cite{BBHP'76} and \cite{Xiao'08}, which have been extensively studied throughout the years. Despite similar effects taking place in either setting, the direct relation has not been established before. The most recent solution \cite{Xiao'08} was motivated by the study of an energy loss of a heavy quark moving through a hot plasma \cite{Mikhailov'03,H3KY'06,Gubser'06,DMMWX'08,CG'08,CST'06,CST'07}. It describes a heavy quark-antiquark pair moving with constant acceleration and connected by the string embedded in the ${\mathrm{AdS}}_{3}$ space. The string worldsheet develops an event horizon and the ensuing physical implications were studied in a number of subsequent works \cite{CCGP'10,CP'11,GIKT'11,GP'11,HKKL'11,CGGP'12,JK'13,CGP'13,Sonner'13,HS'15,Tasgin'19}.
The first string solution \cite{BBHP'76} was derived $45$ years ago and describes a motion of the string with massive ends in $2$d Minkowski spacetime. A particular example of this solution is called a yo-yo string and it corresponds to the case of vanishing masses at the ends of the string or, equivalently, to the limit of strong tension:
\eqlb{eq:yo-yo}{\textrm{yo-yo limit:}\quad m\rightarrow 0 \quad \textrm{or} \quad \gamma \rightarrow \infty.}
In \cite{FG'14} the motivation to study yo-yo strings was the energy loss of gluons and light quarks \cite{GGPR'08,HIM'08,CJK'09,CJKY'09,FNG'13,Ficnar'12}. The yo-yo string was discussed later in the context of the AdS/BCFT correspondence in \cite{AM'15}, but the relation to the AdS solution \cite{Xiao'08} was not noticed. This is probably due to the fact that the relation appears in a slightly more complicated limit than just (\ref{eq:yo-yo}). Recently the general solution \cite{BBHP'76} was studied under the name of the periodic zigzag solution \cite{Dubovsky'18,DD1'20,DD2'20}. It was observed that the zigzag solution exhibits some properties of a quantum black hole \cite{Dubovsky'18} and in the present paper we are going to provide further classical evidence to support this claim.
%Another interesting result of \cite{DD2'20} is the proof of classical integrability for the non-periodic zigzag model.

The paper is organized as follows. In sec. \ref{sec:gauge1} we review the original solution \cite{BBHP'76} for the string with massive ends.
Motivated by \cite{BK'79} in sec. \ref{sec:gauge2} we rewrite the same solution in the conformal gauge, where the string equations of motion look like simple wave equations. Even though we are working with $2$ spacetime dimensions, after choosing the conformal gauge the solution is unique due the boundary conditions. This rewriting of the known solution proves to be very insightful. First, we discover a form of $\mathrm{SL}(2,R)$ symmetry (\ref{eq:g_tr}), which governs the periodicity properties of the full solution (without taking the yo-yo limit). Second, the change of coordinates on the worldsheet of the string is reminiscent of the coordinate change near the horizon. The latter statement can be unfolded as follows. The solution in the timelike gauge is defined on the infinite set of finite time intervals of the form $\bsq{\brc{2n-1}\tau_0,\brc{2n+1}\tau_0}$, $n\in \mathbb{Z}$. At the beginning and at the end of each interval the string is collapsed to a point. When we go to the conformal gauge, each one of these finite time intervals is mapped to the infinite time interval in the corresponding coordinate system and thus the folding points are removed to infinity.

In sec. \ref{sec:Polyakov} this black hole like behavior is further explained by introducing two new coordinate systems on the worldsheet related to the accelerated and stationary observers. In the first coordinate system the time "$\eta$" is the proper time of the accelerated observer sitting at the end of the string. One could think about these coordinates as a string analog of Rindler coordinates. Indeed, if we take the limit of a large string length and zoom in on one of the endpoints, we would recover the original Rindler coordinates (see \cite{Aminov'21} for the details). In the second coordinate system the time "$t$" is the proper time of the stationary observer sitting in the middle of the string. In both coordinate systems the horizon is seen in the induced metric on the worldsheet of the string. In coordinates $\brc{t,\sigma}$ the induced metric is
\eqlb{eq:intro_metric1}{\rmd s_{ws}^2=-\frac{\rmd t^2}{\brc{1-t^2/t_0^2}^2} +\alpha^2\, t_0^2\, \rmd \sigma^2, \quad -t_0\leq t \leq t_0,}
and the horizon $t=t_0\equiv\tau_0$ is situated at the folding points. In these coordinates the solution itself is somewhat reminiscent of the short and long string solutions of the $\mathrm{SL}(2,R)$ WZW model \cite{VLS'98,LS'98,MO'00}.
%Since both short and long string solutions have one more parameter than the periodic zigzag solution, one should introduce a limit to fully connect them.
From this point of view the existence of black holes on the worldsheets of strings was discovered long time ago \cite{ARSS'91,Witten'91}.

In sec. \ref{sec:correspondence} we establish the relation between two classical solutions \cite{BBHP'76} and \cite{Xiao'08} . The main idea is to identify the worldsheet coordinates from both solutions based on the form of the induced metrics. Indeed, the induced metric in the second solution \cite{Xiao'08}
\eq{\rmd s_{ws}^2= -\brc{1-\frac{r^2}{b^2}}\rmd\theta^2 + \frac{\rmd\, r^2}{1-r^2/b^2}}
is equivalent to (\ref{eq:intro_metric1}) upon a Weyl transformation and the following identification of coordinates:
\eq{t=r,\quad \sigma = \frac{\theta}{\alpha\, b},}
where we should also assume $b=t_0$. However, the time coordinate $\theta$ runs from $-\infty$ to $+\infty$, whereas $\sigma$ only belongs to the interval $\bsq{0,\pi}$. This discrepancy is resolved by taking the limit $\alpha\rightarrow +\infty$ and restricting to the positive time interval $\theta\geq 0$. As it was briefly mentioned earlier, this limit is a bit more complicated than the yo-yo limit (\ref{eq:yo-yo}) and can be written in terms of worldsheet coordinates $\brc{t,\sigma}$ and spacetime Minkowski coordinates $\brc{X^0, X^1}$ as
\eq{\quad \sigma=\frac{\theta}{\alpha\, b},\quad X^0 = \frac{2\,m}{\gamma} \, Z_{+}^0, \quad X^1-x_0 = \frac{2\,m}{\gamma}\, Z_{+}^1,
\quad \frac{\gamma}{m} \rightarrow +\infty,}
where
\eq{\sinh\brc{\alpha\,\pi}=\frac{\gamma}{m} \, t_0}
and $\brc{-2\, x_0}$ is the maximum length of the string \cite{BBHP'76} in the Minkowski spacetime. The end result is that both classical solutions describe the embedding of the same string worldsheet in two different spacetimes: one is $2$d Minkowski spacetime and another one is $3$d AdS spacetime. The transformation between coordinates in Minkowski and coordinates on the boundary of AdS will be written in the form of the special conformal transformation similar to the ones studied earlier in \cite{Ryang'15}.

\section{Periodic zigzag solution in the timelike gauge}
\label{sec:gauge1}
Following \cite{BBHP'76}, we consider a pair of point masses joined by a massless string:
\eqlb{eq:mString}{S=\int_{\tau_1}^{\tau_2}\rmd\tau \brc{-m \sqrt{-\brc{\partial_{\tau}X\brc{\tau,0}}^2}-m \sqrt{-\brc{\partial_{\tau}X\brc{\tau,\pi}}^2}-\gamma\int_{0}^{\pi}\rmd\sigma \,\sqrt{-G}}.}
The string is described by the usual Nambu-Goto action \cite{Nambu'70,Goto'71} with
\eq{G=\brc{\partial_{\tau}X}^2\brc{\partial_{\sigma}X}^2- \brc{\partial_{\tau}X^{\mu}\partial_{\sigma}X_{\mu}}^2}
and a $D$-component field $X^{\mu}\brc{\tau,\sigma}$. There are two types of equations of motion:
string equations at $0<\sigma<\pi$ and boundary equations at $\sigma=0,\pi$. The conserved total momentum is a sum of point masses momenta $p^{\mu}\brc{\tau,0},\,p^{\mu}\brc{\tau,\pi}$ and the integral of the string momentum density $\gamma \,K^{\mu}\brc{\tau,\sigma}$. Definitions we are going to use are:
\eq{p^{\mu}\brc{\tau,\sigma}=\frac{m\,\partial_{\tau}X^{\mu}}{\sqrt{-\brc{\partial_{\tau}X}^2}},\quad
K^{\mu}\brc{\tau,\sigma}=\frac1{\sqrt{-G}}\brc{\partial_{\tau}X^{\mu}\brc{\partial_{\sigma}X}^2- \partial_{\sigma}X^{\mu}\brc{\partial_{\tau}X^{\nu}\partial_{\sigma}X_{\nu}}},}
\eq{N^{\mu}\brc{\tau,\sigma}=\frac1{\sqrt{-G}}\brc{\partial_{\sigma}X^{\mu}\brc{\partial_{\tau}X}^2- \partial_{\tau}X^{\mu}\brc{\partial_{\tau}X^{\nu}\partial_{\sigma}X_{\nu}}}.}
Then classical equations of motion can be written as
\eqlb{eq:eqS}{0<\sigma<\pi:\quad\partial_{\tau}K^{\mu}+\partial_{\sigma} N^{\mu}=0,}
\eqlb{eq:eqb0}{\partial_{\tau} p^{\mu}\brc{\tau,0}+\gamma \, N^{\mu}\brc{\tau,0}=0,}
\eqlb{eq:eqbp}{\partial_{\tau} p^{\mu}\brc{\tau,\pi}-\gamma \, N^{\mu}\brc{\tau,\pi}=0,}
and the total momentum is
\eqlb{eq:P_total}{P^{\mu}=p^{\mu}\brc{0}+p^{\mu}\brc{\pi}+\gamma \int_{0}^{\pi}\rmd\sigma K^{\mu}.}

\begin{figure}
\begin{center}
\includegraphics[width=0.6\linewidth]{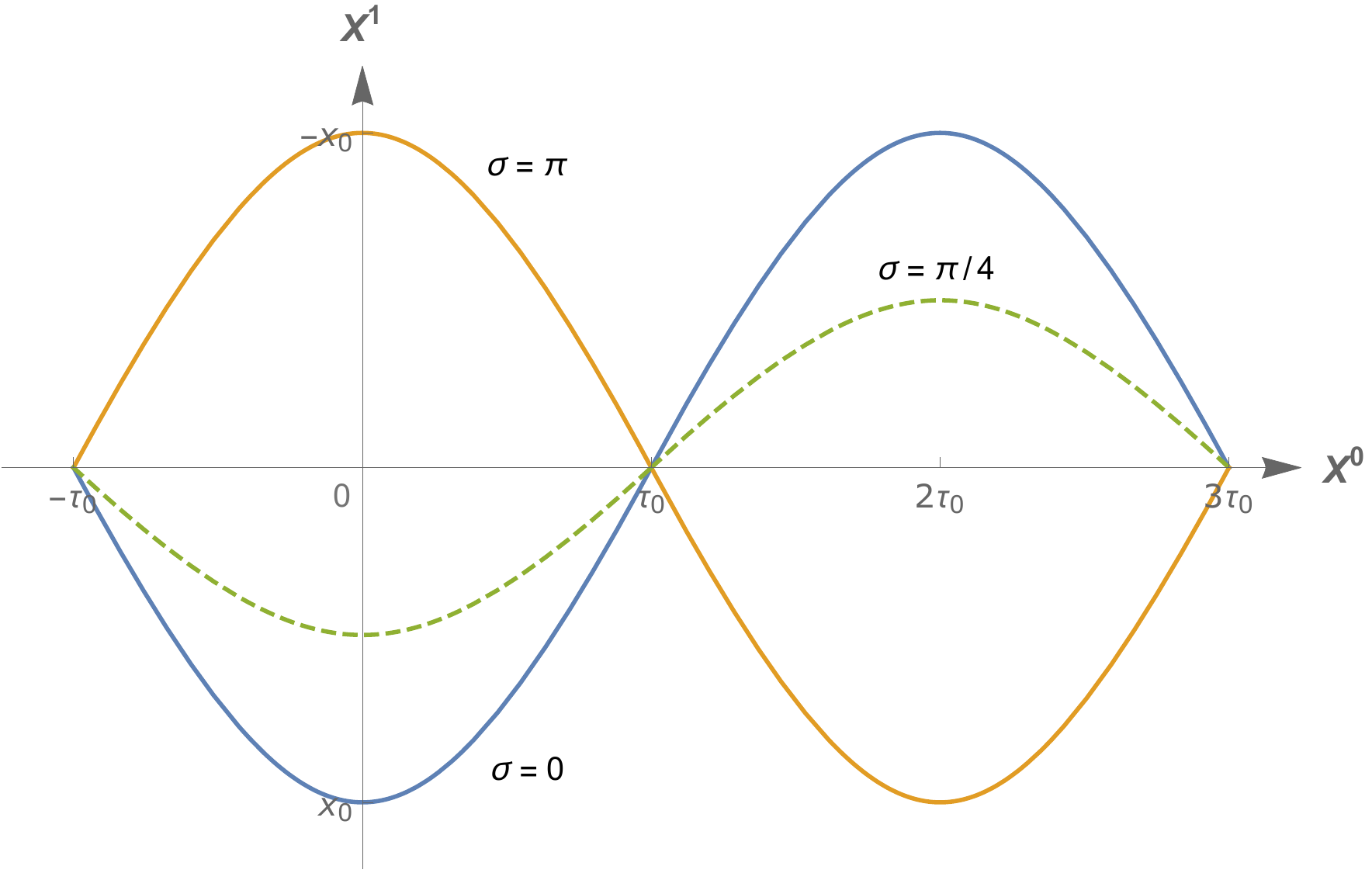}
\caption{Periodic zigzag solution, $-\tau_0\leq \tau\leq 3 \tau_0$}
\label{figure:ZZ_plot}
\end{center}
\end{figure}

Before solving the equations of motion, we need to fix some gauge choices. First, we choose the timelike gauge $X^{0}=\tau$. Second, we can use the $\sigma$ reparametrization invariance to make $\partial_{\sigma} X^{1}$ independent of $\sigma$ in the interval $\brc{0,\pi}$ \cite{BBHP'76}:
\eqlb{eq:gx1}{X^{1}\brc{\tau,\sigma}=\frac{\sigma}{\pi}\brc{X^{1}\brc{\tau,\pi}-X^{1}\brc{\tau,0}}+X^{1}\brc{\tau,0}.}
The periodic zigzag solution describes a $2$d motion, so that only coordinates $X^0$ and $X^1$ are involved. In that case the Hamiltonian is particularly simple:
\eq{H^{2\textrm{d}}=\sqrt{p\brc{0}^2+m^2}+\sqrt{p\brc{\pi}^2+m^2}+\gamma \abs{X^1\brc{\pi}-X^1\brc{0}}.}
In the rest frame we can choose $X^1\brc{\tau,0}=-X^1\brc{\tau,\pi}$ and the solution derived in \cite{BBHP'76} is
\eq{X^1\brc{\tau,\sigma}=\brc{1-\dfrac{2\,\sigma}{\pi}}X^1\brc{\tau,0}}
with the motion of the endpoints described by
\eqlb{eq:x1_sol}{X^1\brc{\tau,0}=\left\{
\begin{matrix}
x_0-\dfrac{m}{\gamma}+\sqrt{\dfrac{m^2}{\gamma^2}+\tau^2}, & \tau\in\brc{-\tau_0,\tau_0},\\
\\
-x_0+\dfrac{m}{\gamma}-\sqrt{\dfrac{m^2}{\gamma^2}+\brc{\tau-2\,\tau_0}^2}, & \tau\in\brc{\tau_0,3\tau_0},
\end{matrix}
\right. }
where we introduced $x_0<0$ and
\eqlb{eq:tau0}{\tau_0^2\equiv x_0^2-\frac{2\,m}{\gamma} x_0.}
The movement is periodic with period $T=4\,\tau_0$. In what follows we are going to be mostly interested in the time interval $\brc{-\tau_0,\tau_0}$. At $\tau=0$ the endpoints are at maximum separation of $2\, x_0$ and start moving towards each other (see Fig. \ref{figure:ZZ_plot}).

\section{Periodic zigzag solution in the conformal gauge}
\label{sec:gauge2}
The goal of this section is to rewrite the $2$d solution in another gauge, where the string equations of motion become simple wave equations. In order to achieve this, the following constraints should be imposed \cite{BK'79}:
\eqlb{eq:g_cons}{\brc{\partial_{\tau}X \pm \partial_{\sigma}X}^2=0,
\quad \brc{\partial^2_{\tau}X \pm \partial_{\tau}\partial_{\sigma}X}^2=\textrm{const}.}
The first constraint leads to the simplified expressions for $G$, $K^{\mu}$ and $N^{\mu}$:
\eqlb{eq:simpW}{G=-\brc{\partial_{\tau}X^{\mu}\partial_{\tau}X_{\mu}}^2, \quad
K^{\mu}=-\partial_{\tau}X^{\mu}\,\frac{\brc{\partial_{\tau}X}^2}{\lvert \brc{\partial_{\tau}X}^2\rvert}, \quad
N^{\mu}=\partial_{\sigma}X^{\mu}\,\frac{\brc{\partial_{\tau}X}^2}{\lvert \brc{\partial_{\tau}X}^2\rvert}.}
Taking into the account that $\brc{\partial_{\tau}X}^2<0$, we can rewrite the equations of motion in the form
\eqlb{eq:eqS}{0<\sigma<\pi:\quad\partial^{2}_{\tau}X^{\mu} - \partial^2_{\sigma} X^{\mu}=0,}
\eqlb{eq:eqb0}{\partial_{\tau} p^{\mu}\brc{\tau,0}=\gamma \, \partial_{\sigma}X^{\mu}\brc{\tau,0},}
\eqlb{eq:eqbp}{\partial_{\tau} p^{\mu}\brc{\tau,\pi}=-\gamma \, \partial_{\sigma}X^{\mu}\brc{\tau,\pi}.}

Using the notations introduced in \cite{BK'79}, we can solve the boundary equation at $\sigma=0$ with the help of a new function $\xi\brc{\tau}$ and a vector $e^{\mu}\brc{\tau}$:
\eqlb{eq:b1}{\left.\partial_{\tau}X^{\mu}\right|_{\sigma=0}=\xi\brc{\tau}\,e^{\mu}\brc{\tau},\quad
\left.\partial_{\sigma}X^{\mu}\right|_{\sigma=0}=\frac{m}{\gamma}\,\dot{e}^{\mu}\brc{\tau}\textrm{sign}\brc{\xi},}
where $e^{\mu}e_{\mu}=-1$ and $\xi^2=m^2\brc{\dot{e}}^2/\gamma^2$. For the $2$d solution we can choose $e^{\mu}$ and $\xi$ to be
\eq{e^{\mu}\brc{\tau}=\brc{\sqrt{1+f\brc{\tau}^2}, f\brc{\tau}}, \quad \xi\brc{\tau}=\frac{m}{\gamma} \frac{\dot{f}\brc{\tau}}{\sqrt{1+f\brc{\tau}^2}}.}
With this choice the second constraint from (\ref{eq:g_cons}) is satisfied and the constant is exactly zero. Therefore, the second constraint does not provide any new information in the $2$d case.

The remaining problem is to find such function $f\brc{\tau}$, that the boundary equation at $\sigma=\pi$ is satisfied. This could be done by demanding that the derivatives of $X^{\mu}$  at $\sigma=\pi$ satisfy similar equations to the ones in (\ref{eq:b1}), except one of the equations should have an additional minus sign on the right hand side and the new functions $\tilde{\xi}$ and $\tilde{e}^{\mu}$ should be used. This results in a peculiar behaviour of a function $f\brc{\tau}$ under the $2\pi$--shifts of the argument:
\eqlb{eq:fpi1}{f\brc{\tau+\pi}+\sqrt{1+f\brc{\tau+\pi}^2}=g\brc{\tau+\pi}-c_1,}
\eqlb{eq:fpi2}{f\brc{\tau-\pi}-\sqrt{1+f\brc{\tau-\pi}^2}=-\frac1{g\brc{\tau+\pi}}-c_2,}
where we introduced two new constants $c_{1,2}$ and a new function $g\brc{\tau}$ with the following properties:
\eqlb{eq:npcond}{c_1<0,\quad c_2>0,\quad c_1< g\brc{\tau} <0,\quad g'\brc{\tau}>0.}

Before we move on with the analysis, lets discuss the change of variables, which occurs when we go from the timelike gauge of the previous section to the conformal gauge in the current one. If we rename the coordinates on the  string worldsheet in the timelike gauge as $\brc{\tilde{\tau},\tilde{\sigma}}$, then the change of coordinates reads
\eqlb{eq:x0_ch}{X^{0}=\tilde{\tau}=\omega^0 +\frac{m}{2\gamma}\left.\brc{f+\sqrt{1+f^2}}\right|_{\brc{\tau+\sigma}} + \frac{m}{2\gamma}\left.\brc{f-\sqrt{1+f^2}}\right|_{\brc{\tau-\sigma}},}
\eqlb{eq:x1_ch}{X^{1}=\brc{1-\frac{2\tilde{\sigma}}{\pi}}\brc{x_0-\frac{m}{\gamma}+\sqrt{\frac{m^2}{\gamma^2}+\tilde{\tau}^2}}=
\omega^1 +\frac{m}{2\gamma}\left.\brc{f+\sqrt{1+f^2}}\right|_{\brc{\tau+\sigma}} - \frac{m}{2\gamma}\left.\brc{f-\sqrt{1+f^2}}\right|_{\brc{\tau-\sigma}},}
where we explicitly used the solution for the time interval of $-\tau_0\leq\tilde{\tau}\leq\tau_0$. Now, the string action is invariant under the reparametrizations of $\tilde{\tau}=\tilde{\tau}\brc{\tau,\sigma}$ and $\tilde{\sigma}=\tilde{\sigma}\brc{\tau,\sigma}$, but the endpoints of the string demand that the reparametrizations of $\tilde{\sigma}$ obey
\eqlb{eq:scond}{\tilde{\sigma}\brc{\tau,0}=0,\quad \tilde{\sigma}\brc{\tau,\pi}=\pi.}
The latter provides us with the consistency check on the behaviour of the function $f\brc{\tau}$ as well as the relations between the various constants. At $\sigma=0$ we have
\eq{\tilde{\tau}\brc{\tau,0}=\omega^0 +\frac{m}{\gamma}f\brc{\tau},}
\eq{\brc{1-\frac{2\tilde{\sigma}\brc{\tau,0}}{\pi}}\brc{x_0-\frac{m}{\gamma}+\sqrt{\frac{m^2}{\gamma^2}+\tilde{\tau}\brc{\tau,0}^2}}=
\omega^1 +\frac{m}{\gamma}\sqrt{1+f\brc{\tau}^2}.}
Thus, the first condition in (\ref{eq:scond}) is satisfied if
\eq{\omega^0=0,\quad \omega^1=x_0-\frac{m}{\gamma}.}
And for the time intervals different from $\brc{-\tau_0, \tau_0}$ the $\omega^0$ constant would be non-zero. Next, we are going to look at $\sigma=\pi$ and use the equations (\ref{eq:fpi1})--(\ref{eq:fpi2}):
\eq{\tilde{\tau}\brc{\tau,\pi}=\frac{m}{2\gamma}\brc{g\brc{\tau+\pi}-c_1-\frac1{g\brc{\tau+\pi}}-c_2},}
\eq{\brc{1-\frac{2\tilde{\sigma}\brc{\tau,\pi}}{\pi}}\brc{x_0-\frac{m}{\gamma}+\sqrt{\frac{m^2}{\gamma^2}+\tilde{\tau}\brc{\tau,\pi}^2}} = \omega^1+\frac{m}{2\gamma}\brc{g\brc{\tau+\pi}-c_1+\frac1{g\brc{\tau+\pi}}+c_2}.}
We conclude that to satisfy the second condition in (\ref{eq:scond}) the following relations should hold:
\eq{c_2=-c_1,\quad c_1=\frac{2\gamma}{m}\,\omega^1=2\brc{\frac{\gamma}{m}x_0-1}.}
This leaves us with the task of determining the function $g\brc{\tau}$. First, we want to rewrite (\ref{eq:fpi1})--(\ref{eq:fpi2}) in terms of the $g$-function alone:
\eqlb{eq:g_tr}{g\brc{\tau+2\pi}=\frac{g\brc{\tau}-c_1}{1+c_1 c_2-c_2\, g\brc{\tau}}.}
As one can see, $g\brc{\tau}$ and $g\brc{\tau+2\pi}$ are connected by the $\textrm{SL}\brc{2,R}$ transformation.
To find a suitable candidate for the solution, it is useful to take a step back and think about the equations of motion. In the timelike gauge we had a sign function involved in the boundary equations, which made the time derivatives of the point particles momenta to be discontinuous. In the conformal gauge, on the other hand, we do not see any discontinuities neither in the boundary equations (\ref{eq:eqb0})--(\ref{eq:eqbp}) nor in the string equations of motion (\ref{eq:eqS}). Moreover, if we look at the change of variables (\ref{eq:x0_ch})--(\ref{eq:x1_ch}), we see that the same points of discontinuity $\tilde{\tau}=\pm\,\tau_0$ make impossible to determine $\tilde{\sigma}\brc{\pm\,\tau_0, \sigma}$ as a function of $\sigma$. These mismatches between the two coordinate systems on the string worldsheet could be resolved, if we map the finite time interval $\brc{-\tau_0,\tau_0}$ in the timelike gauge to the infinite one in the conformal gauge. This can also be seen in (\ref{eq:npcond}), where the $g$-function is described as being bounded from above and below. A good example of a real-valued bounded function, which also has a positive derivative everywhere, is a hyperbolic tangent. Now we are ready to solve (\ref{eq:g_tr}).
By implementing the addition formula for the hyperbolic tangent
\eq{\tanh\brc{\alpha+\beta}=\frac{\tanh\brc{\alpha}+\tanh\brc{\beta}}{1+\tanh\brc{\alpha}\tanh\brc{\beta}}}
and assuming the general form of the $g$-function to be $g\brc{\tau}=a\,\tanh\brc{\alpha\,\tau}+b$, we can solve for the constants $a$, $\alpha$ and $b$ to find the solution:
\eq{g\brc{\tau}=\frac{1}{2}\brc{\sqrt{c_1^2-4}\,\tanh\brc{\alpha\,\tau}+c_1},\quad
\tanh\brc{\alpha\, \pi}=\sqrt{1-\frac{4}{c_1^2}}.}
Here we would like to emphasize, that we were simply solving the equation (\ref{eq:g_tr}) and not using any information about the map between the time intervals in different gauges. So, the last consistency check is to verify that the time interval $-\tau_0\leq\tilde{\tau}\leq\tau_0$ in the timelike gauge is mapped to the infinite time interval in the conformal gauge at any value of $\sigma$ on the string.
Using the definition of $\tilde{\tau}\brc{\tau,\sigma}$ from (\ref{eq:x0_ch}) and taking the limits $\tau\rightarrow\pm\infty$, we get
\eq{\tilde{\tau}\brc{\pm\infty,\sigma}=\frac{m}{\gamma}f\brc{\pm\infty}.}
We can already see, that the dependence on $\sigma$ disappears in these limits. Rewriting the function $f\brc{\tau}$ in terms of the $g$-function
\eq{f\brc{\tau}=\frac{1}{2}\brc{g\brc{\tau}-c_1-\frac{1}{g\brc{\tau}-c_1}}}
and calculating the limits for $g\brc{\tau}$
\eq{g\brc{\pm\infty}=\frac{1}{2}\brc{c_1\pm \sqrt{c_1^2-4}},}
we arrive at the positive conclusion
\eq{\tilde{\tau}\brc{\pm\infty,\sigma}=\pm \frac{m}{2\gamma}\sqrt{c_1^2-4}=\pm\tau_0.}

To summarize, we recall that the $2$d solution in the timelike gauge is defined on the infinite set of the finite time intervals of the form $\bsq{\brc{2n-1}\tau_0,\brc{2n+1}\tau_0}$, $n\in \mathbb{Z}$. When we go to the conformal gauge, each one of these finite time intervals is mapped to the infinite time interval in the corresponding coordinate system. In other words, each finite patch of worldsheet coordinates in the timelike gauge, described by
\eq{\brc{2n-1}\tau_0\leq\tilde{\tau}\leq\brc{2n+1}\tau_0 \quad \textrm{and} \quad 0\leq\tilde{\sigma}\leq\pi,}
is mapped to the corresponding infinite patch in the conformal gauge:
\eq{-\infty\leq\tau_{\brc{n}}\leq+\infty \quad \textrm{and} \quad 0\leq\sigma\leq\pi.}
This kind of behavior is indeed reminiscent of the black hole solution and the coordinate change near the horizon \cite{Dubovsky'18}.

\section{Polyakov action and the worldsheet metric}
\label{sec:Polyakov}
In this section we are going to introduce two new coordinate systems on the worldsheet of the string related to the accelerated and stationary observers. This will help us further understand the black hole like properties of the periodic zigzag solution. In the first coordinate system the time "$\eta$" is the proper time of the accelerated observer sitting at the end of the string. One could think about these coordinates as a string analog of Rindler coordinates. In the second coordinate system the time "$t$" is the proper time of the stationary observer sitting in the middle of the string. In both coordinate systems the horizon is seen in the induced metric on the worldsheet.

\subsection{Metric for the accelerated observer}
In this subsection we want to derive an analog of Rindler coordinates. First, we rewrite the string action in terms of the worldsheet metric $h_{\alpha\beta}$:
\eq{S_{\textrm{string}}=-\frac{\gamma}{2}\int\rmd\tau\,\rmd\sigma\, \sqrt{-h}\,h^{\alpha\beta} \brc{\partial_{\alpha}X^{\mu}\partial_{\beta}X_{\mu}}.}
In general, string equations of motion then take the following form:
\eq{\frac{1}{\sqrt{-h}}\partial_{\alpha}\brc{\sqrt{-h}\,h^{\alpha\beta} \partial_{\beta}X^{\mu}}=0.}
Since the Polyakov action is Weyl invariant, we conclude that the metric in coordinates $\brc{\tau,\sigma}$ is flat:
\eq{\rmd s^2=h_{\alpha\beta}\, \rmd \sigma^{\alpha}\, \rmd \sigma^{\beta} =-\rmd \tau^2 +\rmd \sigma^2.}
However, in the limit $\tau \rightarrow \pm \infty$ the metric exhibits some $\frac{0}{0}$ ambiguity, as seen from (\ref{eq:simpW}) and the fact that $\brc{\partial_\tau X}^2 \rightarrow 0$ in this limit.

In order to find the string analog of Rindler coordinates, we introduce the following gauge fixing in terms of coordinates $\brc{\eta, \sigma}$:
\eq{\brc{\partial_{\eta}X^{\mu}\partial_{\sigma}X_{\mu}}=0,\quad \brc{\partial_{\sigma}X}^2=-\rho\brc{\eta}^2\brc{\partial_{\eta}X}^2,\quad
\left.\brc{\partial_{\eta}X}^2\right|_{\sigma=0,\pi}=-1,}
where the last condition indicates that $\eta$ is a proper time for the endpoints of the string and the function $\rho\brc{\eta}$ is fixed by the equations of motion. In these coordinates the string equations of motion take the following form:
\eqlb{eq:eom_eta}{\partial_{\eta}\brc{\abs{\rho\brc{\eta}}\partial_{\eta}X^{\mu}}- \partial_{\sigma}\brc{\frac1{\abs{\rho\brc{\eta}}}\partial_{\sigma}X^{\mu}}=0}
and the metric is
\eqlb{eq:metric_R}{\rmd s^2 =-\frac1{\rho\brc{\eta}^2} \rmd \eta^2 + \rmd \sigma^2.}
The $2$d solution can be obtained with the help of the coordinate change
\eq{g\brc{\tau}=c_1+\rme^{\frac\gamma{m}\eta},}
which gives
\eqlb{eq:x0_eta}{X^{0}=\frac{m}{\gamma}\,\frac{\eta_0 \sinh\brc{\frac{\gamma}{m}\eta_0}\sinh\brc{\frac{\gamma}{m}\eta}}
{\eta_0 \sinh\brc{\frac{\gamma}{m}\eta_0}-\pi\rho\brc{\eta} \sinh\brc{\alpha\, \sigma} \sinh\brc{\alpha\brc{\pi-\sigma}}},}
\eqlb{eq:x1_eta}{X^{1}=x_0-\frac{m}{\gamma}+\frac{m}{\gamma}\,\frac{\eta_0 \sinh\brc{\frac{\gamma}{m}\eta_0}\cosh\brc{\frac{\gamma}{m}\eta}
 +\pi \rho\brc{\eta} \sinh\brc{\alpha\,\sigma}^2}
{\eta_0 \sinh\brc{\frac{\gamma}{m}\eta_0}-\pi\rho\brc{\eta} \sinh\brc{\alpha\, \sigma} \sinh\brc{\alpha\brc{\pi-\sigma}}}.}
And for the function $\rho$ we have:
\eq{\rho\brc{\eta}=-\frac{4\, \eta_0}{\pi}\, \frac{\sinh\brc{\frac{\gamma}{2m}\brc{\eta+\eta_0}}\sinh\brc{\frac{\gamma}{2m}\brc{\eta-\eta_0}}}
{\sinh\brc{\frac{\gamma}{m}\eta_0}}.}
Here we also introduced the constant $\eta_0$, which corresponds to $\tau_0$ in $\brc{\tilde{\tau},\tilde{\sigma}}$ coordinates:
\eq{\tau_0=\frac{m}{\gamma}\sinh\brc{\frac{\gamma}{m}\eta_0},\quad X^1\brc{\pm \eta_0,\sigma}=0.}
Another useful relation between the constants is
\eq{\pi\,\alpha=\frac{\gamma}{m} \eta_0.}
Now we can see, that the solution in coordinates $\brc{\eta,\sigma}$ is very similar to the solution described in (\ref{eq:x1_sol}). To be more precise, we only wrote the solution for the time interval $\brc{-\eta_0, \eta_0}$ and the generalization to other time intervals is straightforward.

\subsection{Metric for the stationary observer}
\label{sec:stat_met}
Now that we determined the string analog of the Rindler metric, it is natural to turn to the stationary observer. The obvious placement for such an observer is the middle of the string at $\tilde{\sigma}=\sigma=\pi/2$. Indeed, at $\sigma=\pi/2$ we have
\eq{X^{0}\brc{\eta, \frac{\pi}2}=\frac{2m}{\gamma} \cosh\brc{\frac{\gamma}{2 m} \eta_0}^2 \tanh\brc{\frac{\gamma}{2 m} \eta},
\quad X^{1}\brc{\eta, \frac{\pi}2}=0.}
Thus, we can introduce the proper time $t$ for the stationary observer as
\eqlb{eq:st_t}{t\brc{\eta}=\frac{2m}{\gamma} \cosh\brc{\frac{\gamma}{2 m} \eta_0}^2 \tanh\brc{\frac{\gamma}{2 m} \eta}.}
Then the gauge fixing in coordinates $\brc{t,\sigma}$ is
\eqlb{eq:gauge_t}{\brc{\partial_{t}X^{\mu}\partial_{\sigma}X_{\mu}}=0,\quad \brc{\partial_{\sigma}X}^2=-r\brc{t}^2 \brc{\partial_{t}X}^2,\quad \left.\brc{\partial_{t}X}^2\right|_{\sigma=\pi/2}=-1,}
where
\eq{r\brc{t}=\frac{\alpha}{t_0}\brc{t_0^2-t^2},\quad t_0\equiv t\brc{\eta_0}.}

Going back to the coordinates $\brc{\tilde{\tau},\tilde{\sigma}}$, we can see that the time coordinate $\tilde{\tau}$ is also a good candidate for the proper time of the stationary observer in the following sense:
\eq{\left.\brc{\partial_{\tilde{\tau}}X}^2\right|_{\tilde{\sigma}=\pi/2}=-1.}
So, there is no surprise that from (\ref{eq:st_t}) we get
\eq{t_0=\frac{m}{\gamma}\sinh\brc{\frac{\gamma}{m}\eta_0}=\tau_0.}
However, the induced metric in coordinates $\brc{\tilde{\tau},\tilde{\sigma}}$ is non-diagonal and cumbersome. In coordinates $\brc{t,\sigma}$, on the other hand, the metric reads
\eqlb{t_metric}{\rmd s^2=-\frac{\rmd t^2}{\brc{1-t^2/t_0^2}^2} +\alpha^2\, t_0^2\, \rmd \sigma^2, \quad -t_0\leq t \leq t_0.}
The latter is directly related to the worldsheet metric, which appears in the exact string solution in the $\textrm{AdS}_3$ space \cite{Xiao'08}.

\section{$2$d/$3$d string correspondence}
In this section we are going to establish the relation between two classical solutions \cite{BBHP'76} and \cite{Xiao'08} . The main idea is to identify the worldsheet coordinates from both solutions based on the form of the induced metrics.
\label{sec:correspondence}
\subsection{The exact string solution in the $\textrm{AdS}_3$ space}
We start by writing down the Nambu-Goto action in the $\textrm{AdS}_3$ space:
\eq{S=-T_0\int\, \rmd t\,\rmd u \sqrt{-G},}
where
\eq{G=\brc{\partial_{t}Y}^2\brc{\partial_{u}Y}^2- \brc{\partial_{t}Y^{\mu}\partial_{u}Y_{\mu}}^2.}
The $3$-component field $Y^{\mu}$ is a vector in the $\textrm{AdS}_3$ spacetime with the metric
\eq{\rmd s^2 = R^2\bsq{-\brc{Y^1}^2 \brc{\rmd Y^0}^2 +\frac{\brc{\rmd Y^1}^2}{\brc{Y^1}^2}+\brc{Y^1}^2 \brc{\rmd Y^2}^2}.}
The solution found in \cite{Xiao'08} describes a motion of the string in the static gauge
\eq{Y^{\mu}=\brc{t,u,y\brc{t,u}}}
and can be written as
\eqlb{eq:Asol}{y=\pm \sqrt{t^2+b^2 -\frac{1}{u^2}}.}
The constant $b$ is the reciprocal of the acceleration of the string and the two branches of the square root a glued together at $y=0$.
Since the solution describes a string with constant proper acceleration, one can go from the AdS space to a generalized Rindler space. The transformation for the positive branch $y>0$ is
\eq{y=\sqrt{b^2 - r^2} \exp\brc{\frac{\alpha}{b}} \cosh\brc{\frac{\theta}{b}},}
\eq{t=\sqrt{b^2 - r^2} \exp\brc{\frac{\alpha}{b}} \sinh\brc{\frac{\theta}{b}},}
\eq{\frac{1}{u}=r\, \exp\brc{\frac{\alpha}{b}}.}
The same can be done for the negative branch $y<0$. The new coordinates then cover the two wedges $\abs{y}\geq \abs{t}$.
Both $\alpha$ and $\theta$ run from $-\infty$ to $+\infty$, but $r$ only covers the interval $\brc{0,b}$.
The spacetime metric can be rewritten in the new coordinates as
\eqlb{eq:gRindler}{\rmd s^2 = \frac{R^2}{r^2}\bsq{\frac{\rmd\, r^2}{1-r^2/b^2}-\brc{1-\frac{r^2}{b^2}}\rmd \theta^2 +\rmd \alpha^2}.}
The solution (\ref{eq:Asol}) corresponds to $\alpha=0$ and can be viewed as a part of a one-sheet hyperboloid:
\eqlb{eq:1shyp}{y^2+r^2-t^2=b^2.}
In what follows we will consider coordinates $\brc{r,\theta}$ as new coordinates on the string worldsheet, which correspond to a different gauge choice for $Y^{\mu}$:
\eq{Y^{\mu}=\brc{t\brc{r,\theta},\frac1{r},y\brc{r,\theta}}.}
This identification between Rindler spacetime coordinates and a particular choice of worldsheet coordinates also appears in the Minkowski setting in \cite{Aminov'21}. The induced metric on the worldsheet is then
\eqlb{eq:h_ind}{\rmd s_{ws}^2= -\brc{1-\frac{r^2}{b^2}}\rmd\theta^2 +
\frac{\rmd\, r^2}{1-r^2/b^2}.}

\subsection{Mapping between the solutions}
First, we observe that the worldsheet metric in (\ref{eq:h_ind}) coincides with the worldsheet metric in (\ref{t_metric}) upon a Weyl transformation and the following identification of coordinates:
\eq{t_{\brc{z}}=r,\quad \sigma = \frac{\theta}{\alpha\, b},}
where $z$ in the $t_{\brc{z}}$ notation stands for the zigzag solution. Here we see the main difference between the two solutions:
$\theta$ runs from $-\infty$ to $+\infty$, whereas $\sigma$ only belongs to the interval $\bsq{0,\pi}$.
To overcome this obstacle one needs to take a limit $\alpha\rightarrow +\infty$ and consider only the positive time interval $\theta\geq 0$ or, equivalently, $t\geq 0$ in the AdS solution. The same conclusion can be reached by noting that the zigzag solution has two constant parameters $x_0$, $\frac{m}{\gamma}$ and one needs to match them to the single parameter $b$ in the AdS solution. The limit $\alpha\rightarrow +\infty$ corresponds to the strong tension limit as it can be seen from the following relations:
\eq{\sinh\brc{\pi\,\alpha}=\frac{\gamma}{m} \,\tau_0 \rightarrow +\infty,\quad \tau_0 \rightarrow b.}
Using (\ref{eq:tau0}), we get the full matching between the parameters:
\eq{b=-x_0>0,\quad \frac{\gamma}{m} \rightarrow +\infty.}
Thus, the reciprocal of the acceleration in the AdS solution corresponds to the half of the maximum string length in the zigzag solution.

\begin{figure}
\begin{center}
\includegraphics[width=1\linewidth]{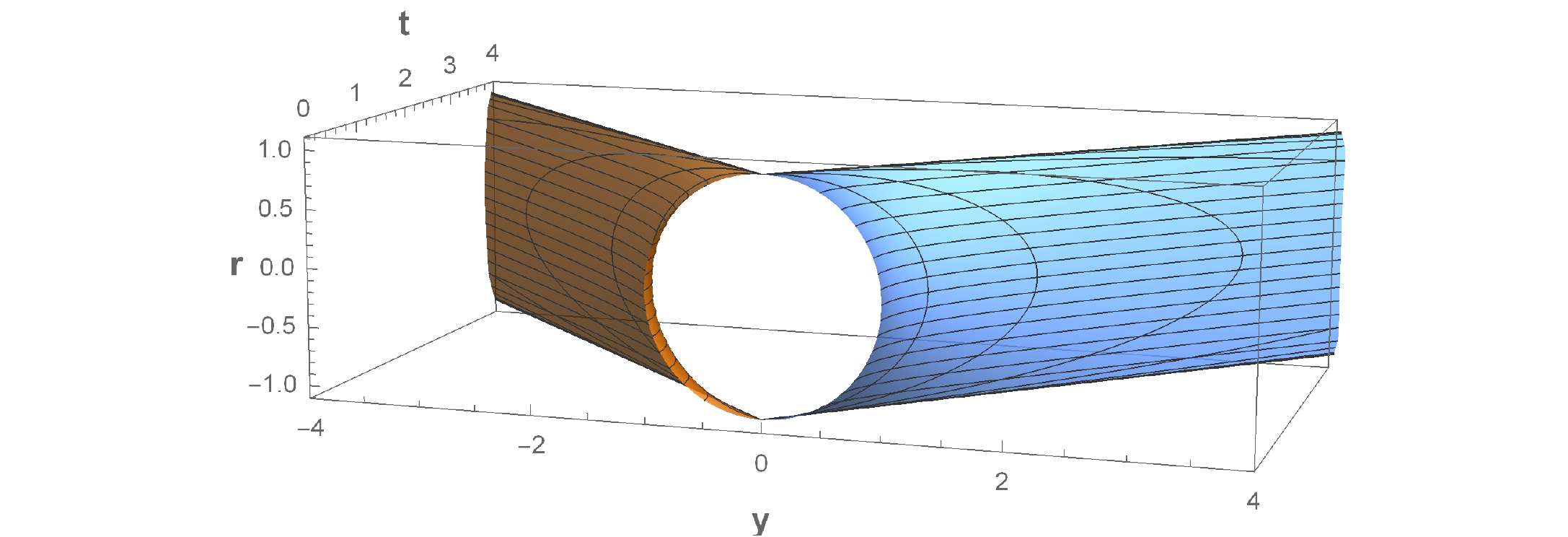}
\caption{Zigzag solution embedded in the AdS space, $b=1$}
\label{figure:3d_AdS}
\end{center}
\end{figure}

To complete the correspondence we want to write down the coordinate transformation between the Minkowski coordinates $X^{\mu}$ in the zigzag solution and coordinates $\brc{t,y}$ on the boundary of AdS. But first, lets introduce a clear geometrical picture of what is going on. The AdS solution can be described as a top part of the hyperboloid (\ref{eq:1shyp}) when it is cut by the plane $r=0$ (the boundary of the AdS space), where the quark and antiquark are moving. The zigzag solution, on the other hand, is described by the part of the hyperboloid with $t\geq 0$ and $-b \leq r \leq b$. Going from $r>0$ to $r=0$ and then to $r<0$ can be explained by the specific identification on the boundary of AdS, which we will discuss in the next subsection \ref{sec:bid_AdS}. Due to the event horizons at $r=\pm b$, we should also cut the hyperboloid by the planes $r=-b$ and $r=b$  (see Figure \ref{figure:3d_AdS}).
The zigzag solution then corresponds to the two branches of the following square root:
\eqlb{eq:pmr}{r=\pm \sqrt{b^2+t^2 - y^2}.}
The positive branch represents $X^0>0$ and the negative one represents $X^0<0$. Since $r$ is the proper time of the stationary observer in the zigzag solution, the $r=0$ plane is exactly the $X^0=0$ line in Minkowski. Thus, the worldlines of the quark and antiquark are mapped to the string embedded in Minkowski space at $X^0=0$. The same is true in reverse. The worldlines of the endpoints of the string in the zigzag solution are mapped to the closed string embedded in the AdS space at $t=0$.

Now we are ready to derive the coordinate transformation $X^{\mu}\brc{t,y}$. First, we consider the positive branch of (\ref{eq:Asol}):
\eq{y=\sqrt{b^2 - r^2} \cosh\brc{\frac{\theta_{+}}{b}},}
\eq{t=\sqrt{b^2 - r^2} \sinh\brc{\frac{\theta_{+}}{b}}.}
Substituting the $\eta\brc{t_{\brc{z}}}$ in (\ref{eq:x0_eta})--(\ref{eq:x1_eta}) and taking the limit $\gamma/ m \rightarrow \infty$ we get
\eqlb{eq:MAdS_sol}{y>0:\quad X^0=\frac{2\,m}{\gamma} \frac{\brc{\pm} b \sqrt{b^2+t^2-y^2}}{\brc{y-t}^2},\quad
X^1-x_0=\frac{2\,m}{\gamma} \frac{b^2-y \brc{y-t}}{\brc{y-t}^2},}
where $\brc{\pm}$ in the expression for $X^0$ corresponds to the two branches of the square root in (\ref{eq:pmr}). Here we see that the strong tension limit is accompanied by the rescaling of the $X^{\mu}$ coordinates near one of the endpoints of the string. In particular, since our limit describes zooming in on the left endpoint with $\sigma=0$, the action introduced in (\ref{eq:mString}) becomes an action for the semi-infinite string with one massive particle at $\sigma=0$. To be more precise, we introduce the following limit in terms of the coordinates $\brc{t_{\brc{z}}\equiv r,\sigma}$ and $\brc{X^0, X^1}$:
\eq{y>0:\quad \sigma=\frac{\theta_{+}}{\alpha\, b},\quad X^0 = \frac{2\,m}{\gamma} \, Z_{+}^0, \quad X^1-x_0 = \frac{2\,m}{\gamma}\, Z_{+}^1,
\quad \frac{\gamma}{m} \rightarrow +\infty.}
Then the functions $Z_{+}^\mu\brc{r,\theta}$ given by (\ref{eq:MAdS_sol}) describe a classical solution to the following string action:
\eq{S=\frac{2\,m^2}{\gamma}\int \rmd r \brc{ -\sqrt{-\brc{\partial_{r}Z_{+}\brc{r,0}}^2} - 2
\int_{0}^{\infty} \rmd \theta \sqrt{-G}},\quad
G=\brc{\partial_{r}Z_{+}}^2\brc{\partial_{\theta}Z_{+}}^2- \brc{\partial_{r}Z_{+}^{\mu}\partial_{\theta}Z^{+}_{\mu}}^2.}
As one can see, the action has an overall scale $m^2/\gamma$ and the equations of motion are independent of the ratio $m/\gamma$.
The coordinate transformation from $\brc{t,y}$ to $\brc{Z_{+}^0, Z_{+}^1}$ can be written in the form of the special conformal transformation:
\eq{Z_{+}^{\mu}=\frac{\hat{y}^{\mu}-\beta^{\mu}\hat{y}^2}{1-2 \beta^{\mu}\hat{y}_{\mu}+\beta^2 \hat{y}^2}}
with
\eq{y>0:\quad \hat{y}^{\mu}= \brc{\frac{b\,r}{b^2+t^2},\frac{t\,y}{b^2+t^2}}, \quad \beta^{\mu}=\brc{0,1}.}

So far we described the "blue" part of the surface with $y>0$ in Figure \ref{figure:3d_AdS}, which corresponds to the worldline of the quark and the attached string in the AdS. To describe the "red" part we consider the negative branch of (\ref{eq:Asol}):
\eq{y=-\sqrt{b^2 - r^2} \cosh\brc{\frac{\theta_{-}}{b}},}
\eq{t=\sqrt{b^2 - r^2} \sinh\brc{\frac{\theta_{-}}{b}},}
and zoom in on the right endpoint of the string with $\sigma = \pi$:
\eq{t_{\brc{z}}=r,\quad \sigma=\pi-\frac{\theta_{-}}{\alpha\,b}, \quad \alpha \rightarrow \infty.}
The corresponding coordinate transformation $X^{\mu}\brc{t,y}$ is then
\eq{y<0:\quad X^0=\frac{2\,m}{\gamma} \frac{\brc{\pm} b \sqrt{b^2+t^2-y^2}}{\brc{y+t}^2},\quad
X^1+x_0=\frac{2\,m}{\gamma} \frac{-b^2+y\brc{y+t}}{\brc{y+t}^2},}
Combining this with (\ref{eq:MAdS_sol}), we get the answer for all allowed values of $y$:
\eqlb{eq:MAdS_full}{\abs{y}\geq t\geq0: \quad X^0=\frac{2\,m}{\gamma} \frac{\brc{\pm} b \sqrt{b^2+t^2-y^2}}{\brc{\abs{y}-t}^2},\quad
X^1=\sgn\brc{y}\brc{-b+ \frac{2\,m}{\gamma} \frac{b^2-\abs{y}\brc{\abs{y}-t}}{\brc{\abs{y}-t}^2}}.}
By introducing local Minkowski variables $Z_{-}^{\mu}$ near the left endpoint of the string with $\sigma=\pi$, we can rewrite (\ref{eq:MAdS_full}) in the form of the special conformal transformation as
\eq{\mu=0,1:\quad Z_{\pm}^{\mu}=(\pm1)^{\mu}\frac{\hat{y}^{\mu}-\beta^{\mu}\hat{y}^2}{1-2 \beta^{\mu}\hat{y}_{\mu}+\beta^2 \hat{y}^2},}
where $\brc{+}$ corresponds to $y>0$, $\brc{-}$ corresponds to $y<0$ and vectors $\hat{y}^{\mu}$, $\beta^{\mu}$ are given by
\eq{\hat{y}^{\mu}= \brc{\frac{b\,r}{b^2+t^2},\frac{t \abs{y}}{b^2+t^2}}, \quad \beta^{\mu}=\brc{0,1}.}
It is helpful to picture the domain in $\brc{y,t}$ variables, which is mapped to the zigzag solution.
As it can be seen from Figure \ref{figure:3d_AdS}, the domain is $t\leq \abs{y}\leq \sqrt{b^2+t^2}$ (shaded regions in Fig. \ref{figure:yt_AdS}).
The boundaries of this region are the worldlines of the quark-antiquark pair $\abs{y}=\sqrt{b^2+t^2}$ and the event horizons $\abs{y}=t$. Since for $r\neq 0$ the hyperboloid covers the domain twice, we have two horizons $Z^0=\pm \infty$ at $y=t$ and at $y=-t$.

The final step in the correspondence is to associate the two lines $\abs{y}=t$ with the worldline of the middle of the string at $\sigma=\pi/2$ in the zigzag solution. To illustrate this point, we are going to project worldlines of different points on the string on the half-planes $y=t\geq0$ and $y=-t\leq0$. As we derived earlier, different points on the string in the zigzag solution are parametrized by the proper time $\theta$ of the accelerated observer in the AdS space. To be more precise, there are two different coordinates $\theta_{+}$ and $\theta_{-}$ describing positive and negative branches of the square root in (\ref{eq:Asol}). In Figure \ref{figure:2d_AdS} we see how the different points of the zigzag string are embedded in the AdS space.
The $\theta_{+}$ coordinate describes the points on the string close to $\sigma=0$  and  $\theta_{-}$ describes the points close to $\sigma=\pi$.
We know from the full zigzag solution, that as both $\theta_{\pm}\rightarrow +\infty$ we are getting closer to the middle of the string. Thus,
by gluing $\theta_{+}$ and $\theta_{-}$ at infinity we are able to continuously move from the left endpoint of the string with $\sigma=0$ to the right one with $\sigma=\pi$. The gluing point $\theta_{\pm}=+\infty$ corresponds to the middle of the string in the zigzag solution.

\subsection{Identifications on the boundary of $\textrm{AdS}_3$}
\label{sec:bid_AdS}
The solution (\ref{eq:1shyp}) is written in terms of the Poincare coordinates $\brc{t,r,y}$. Usually, one considers $r>0$ and in the limit $r\rightarrow +0$ one approaches the boundary of AdS spacetime. However, the solution itself allows for both positive and negative values of $r$ as well as $r=0$. This can be explained by adding the boundary $r\rightarrow \pm 0$ and introducing specific identifications between different points on the boundary. To figure out the exact form of these identifications, we consider global coordinates $\brc{\hat{\tau}, \hat{\rho}, \hat{\theta}}$ on the $\textrm{AdS}_3$ spacetime:
\eq{\rmd s^2= R^2\brc{-\cosh^2\hat{\rho}\,\rmd \hat{\tau}^2 + \rmd \hat{\rho}^2+\sinh^2\hat{\rho}\,\rmd \hat{\theta}^2},}
where $\hat{\rho}\geq 0$, $\hat{\tau}\in \bsq{0,2\pi}$ and $\hat{\theta}\in \bsq{0,2\pi}$. The relation to the Poincare coordinates can be written in the following form:
\eq{
\begin{array}{lcl}
  R\, \cosh\hat{\rho}\,\cos\hat{\tau} & = & r/2 +\brc{R^2+y^2-t^2}/\brc{2\,r}, \\
  \cosh\hat{\rho}\,\sin\hat{\tau} & = & t/r ,\\
  \sinh\hat{\rho}\,\sin\hat{\theta} & = & y/r ,\\
  R\, \sinh\hat{\rho}\,\cos\hat{\theta} & = & r/2 -\brc{R^2-y^2+t^2}/\brc{2\,r} .
\end{array}
}
We can see, that Poincare coordinates with $r>0$ cover only "half" of the AdS spacetime. Changing the sign of $r$ corresponds to a simultaneous shift of $\hat{\tau}$ and $\hat{\theta}$ by $\pi$:
\eq{r\rightarrow -r :\quad \hat{\tau}\rightarrow\hat{\tau}+\pi \quad \textrm{and} \quad \hat{\theta}\rightarrow\hat{\theta}+\pi.}
Thus, two limits $r\rightarrow +0$ and $r\rightarrow -0$ describe two different regions on the AdS boundary. To continuously go from $r>0$ to $r<0$, one needs to identify these boundary regions. In our particular example we consider $t\geq 0$, which leaves us with two cases: $t\geq 0$, $y\geq 0$ and $t\geq 0$, $y\leq 0$. In each case $r>0$ and $r<0$ cover different quadrants in $\brc{\hat{\tau}, \hat{\theta}}$ coordinates (see Figure \ref{figure:global_AdS}). In the limit $r\rightarrow \pm 0$ we identify the points on the boundary according to the following rule (see Figure \ref{figure:bry_AdS}):
\eq{\brc{\hat{\tau},\hat{\theta}}\longleftrightarrow \brc{\hat{\tau}+\pi,  \hat{\theta}+\pi}.}
In this way we get an unconventional Poincare patch with $t\geq 0$ and $-\infty<r,y<+\infty$, which still covers "half" of the $\textrm{AdS}_3$. This patch can be visualized in global coordinates $\brc{\hat{\tau}, \hat{\rho}, \hat{\theta}}$ as in Figure \ref{figure:u_Poincare}.

\section{Conclusion}
In this paper we introduced a new relation between two classical string solutions \cite{BBHP'76} and \cite{Xiao'08}. The first solution describes two point masses joined by a massless string in the $2$d Minkowski spacetime and the second solution describes a quark-antiquark pair connected by a string embedded in the $3$d AdS spacetime. As such, the relation takes place in the lowest possible number of spacetime dimensions and requires a particular form of identification on the boundary of $\textrm{AdS}_3$. It was shown that the latter identification leads to an alternative definition of the Poincare patch. The following physical interpretation of the established relation could be given. The first solution \cite{BBHP'76} provides a model for the motion of two quarks subjected to a strong interaction. The second solution is conjectured to describe a motion of the quark-antiquark pair moving in the external electric field. The external electric field not only balances out the attracting force between the quark and antiquark, but also accelerates them with constant proper acceleration in different directions. If the physical interpretation of the second solution \cite{Xiao'08} proves to be correct, the result of the present paper suggests a non-trivial relation between these two different physical systems. Since we covered classical aspects of the said relation, a straightforward continuation would be to introduce a quantum version of the correspondence. Another interesting direction is to study physical effects like the Unruh effect on both sides of the correspondence. In the case of the second solution the Unruh effect was discussed in the original paper \cite{Xiao'08} and the corresponding temperature was compared to the Hawking temperature determined by the behavior of the metric near the horizon. In the case of the first solution it is possible to derive an analog of the Unruh effect by adding small transversal excitations and taking the limit of a large string length. This will be discussed in a subsequent paper \cite{Aminov'21}.

\section*{Acknowledgements}
The author would like to thank Martin Roček and Zohar Komargodski for many valuable discussions and comments. The work was supported by the "BASIS" Foundation grant 18-1-1-50-3 and in part by RFBR grant 19-51-18006-Bolg\_a.

\appendix
\section{Additional figures}

\begin{figure}[!h]
\begin{center}
\includegraphics[width=0.7\linewidth]{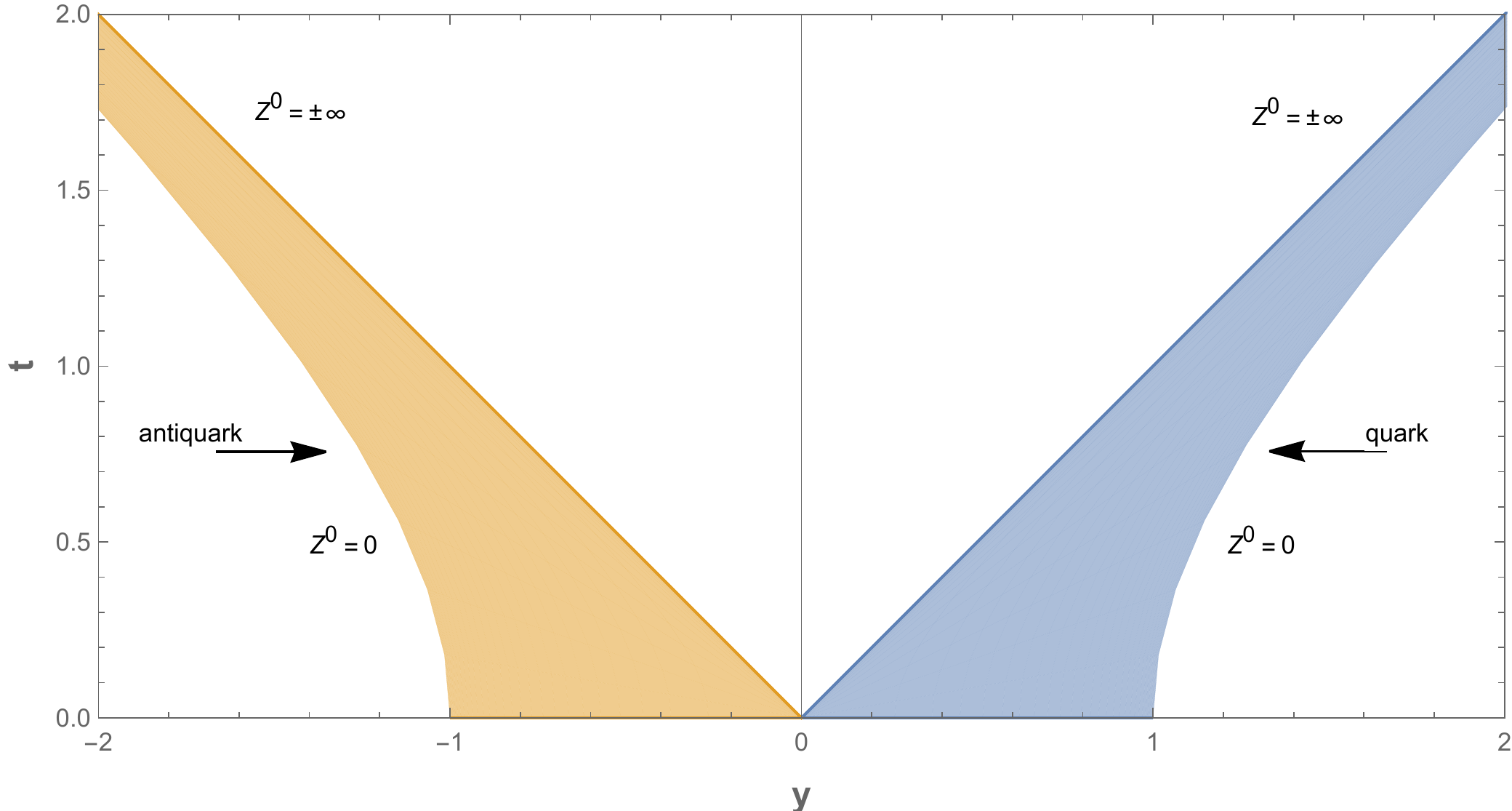}
\caption{$\brc{y,t}$ domain, $b=1$}
\label{figure:yt_AdS}
\end{center}
\end{figure}

\begin{figure}[!h]
\begin{center}
\includegraphics[width=0.7\linewidth]{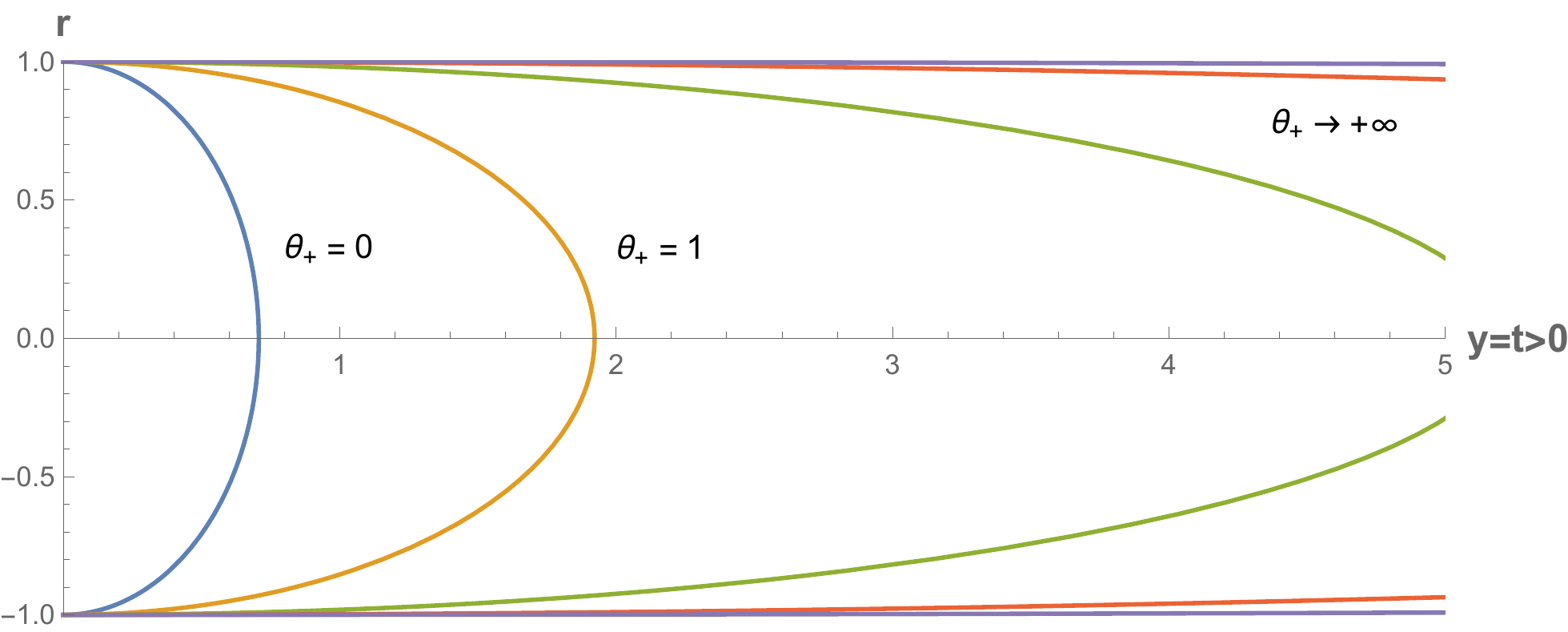}
\includegraphics[width=0.7\linewidth]{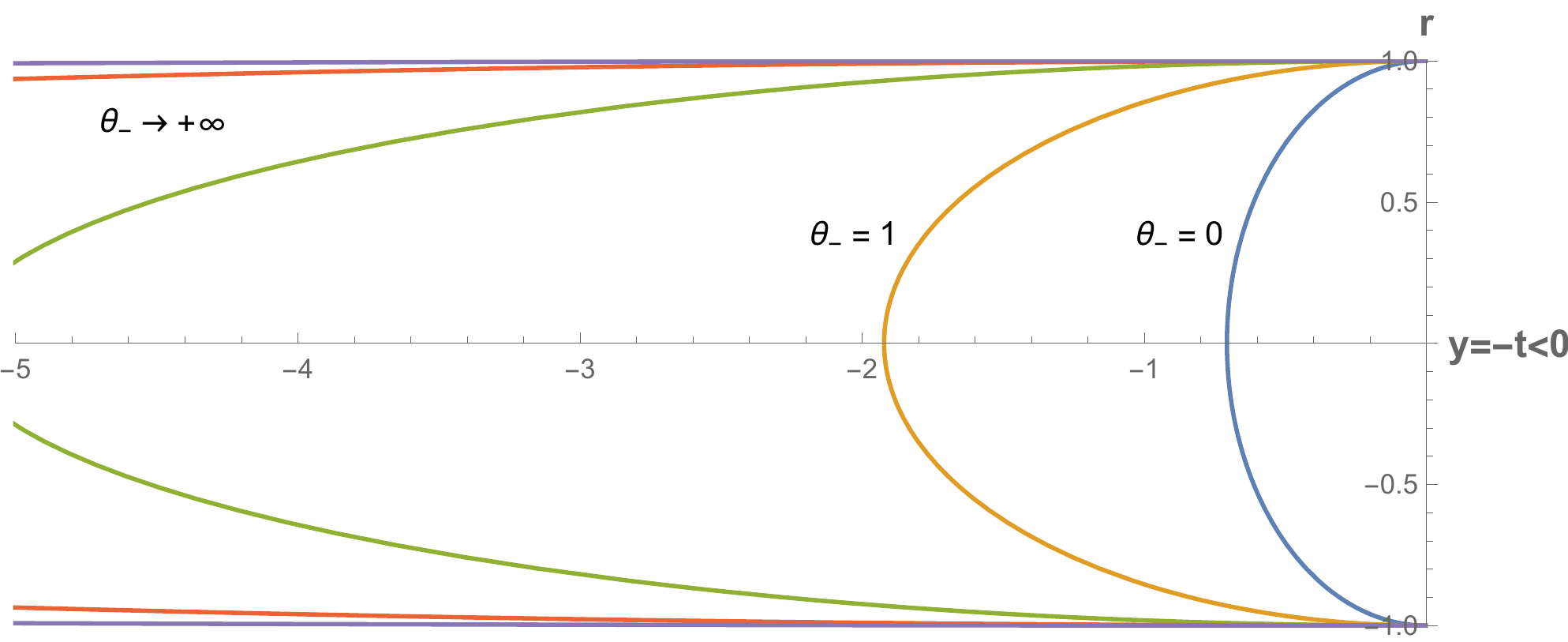}
\caption{Worldlines of different points on the zigzag string, $y=\pm t$ and $b=1$}
\label{figure:2d_AdS}
\end{center}
\end{figure}

\begin{figure}[!h]
\begin{center}
\includegraphics[width=0.4\linewidth]{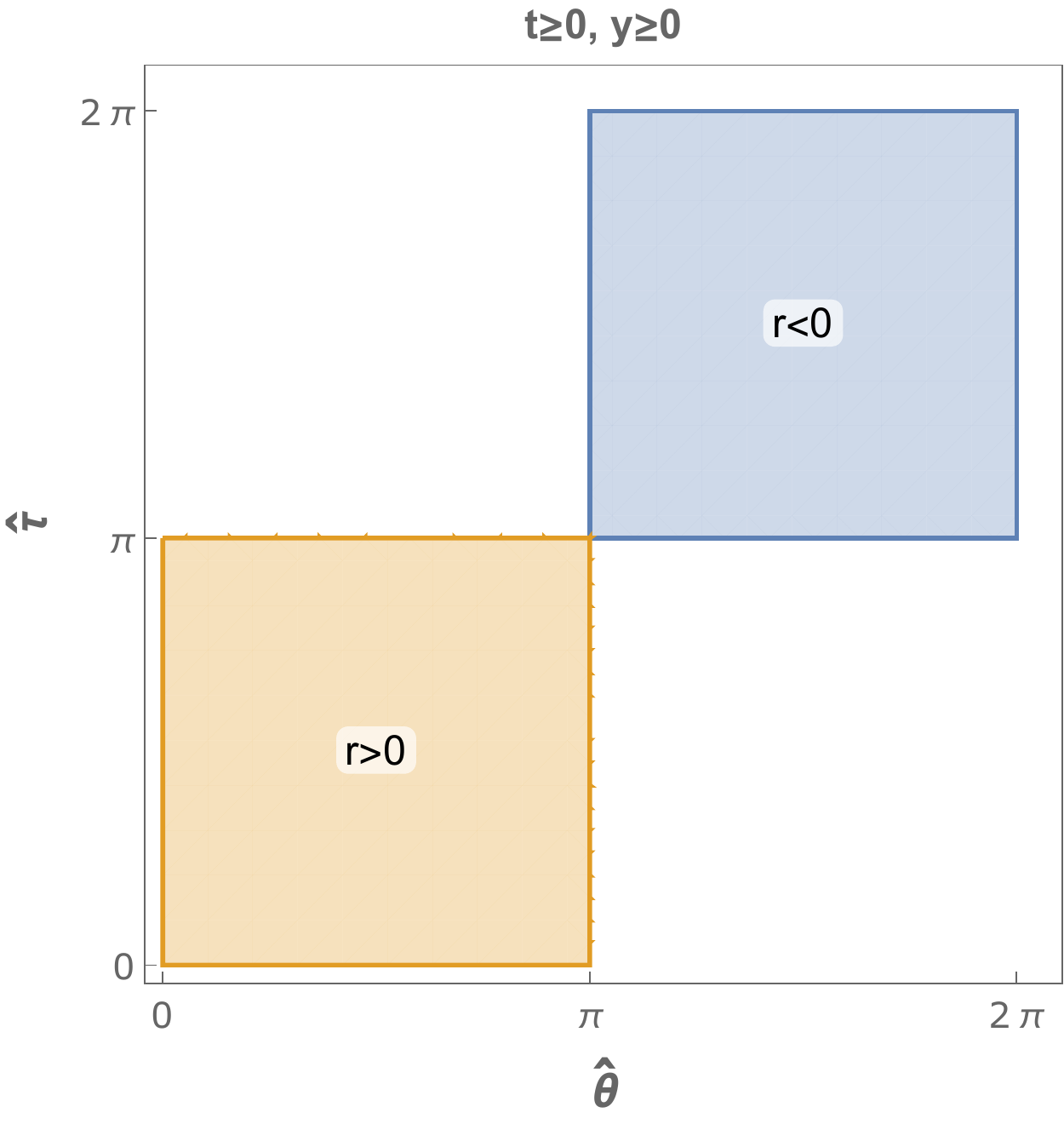}\quad
\includegraphics[width=0.4\linewidth]{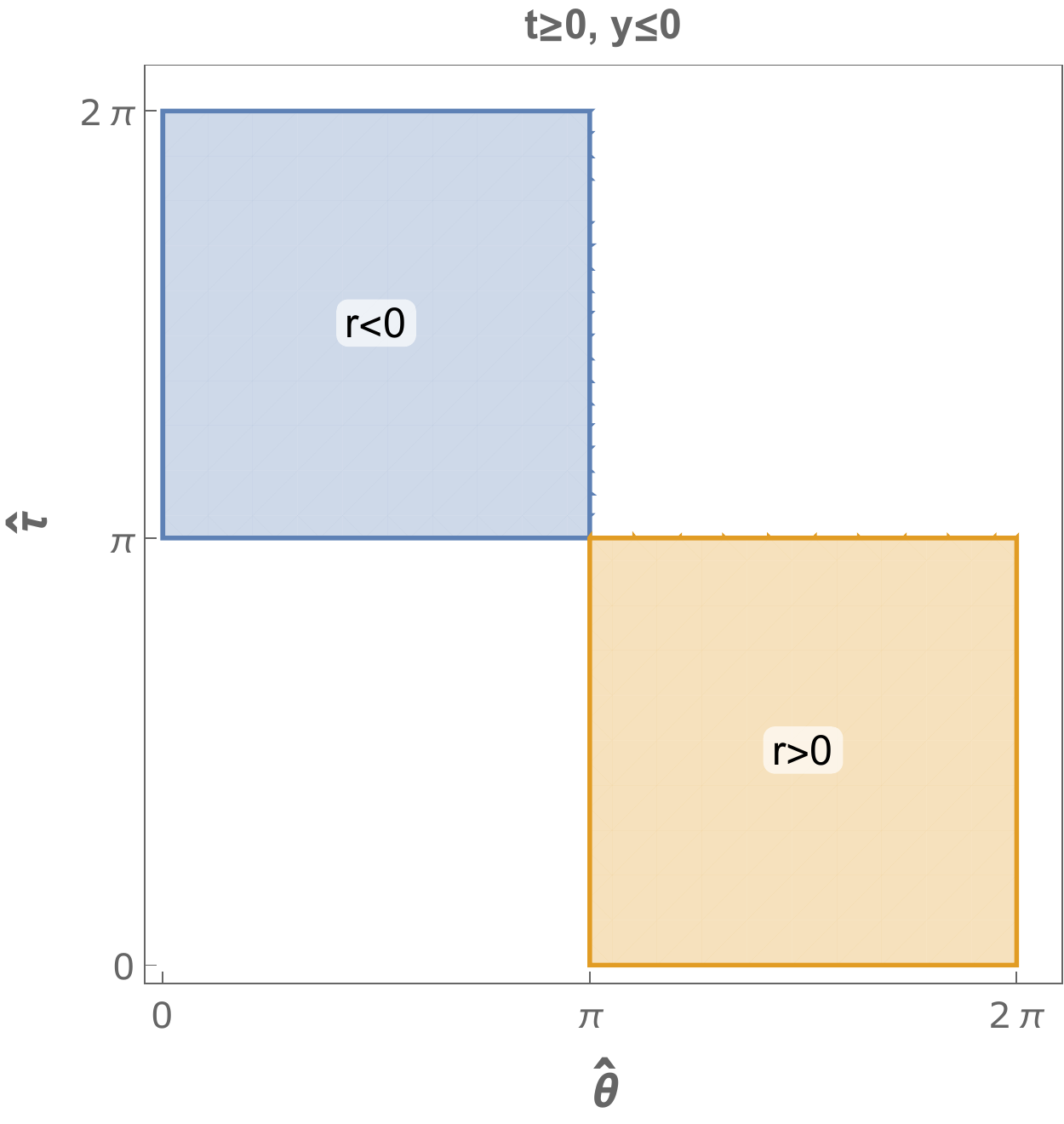}
\caption{Regions with $r>0$ and $r<0$ in $\brc{\hat{\tau}, \hat{\theta}}$ coordinates}
\label{figure:global_AdS}
\end{center}
\end{figure}

\begin{figure}[!h]
\begin{center}
\includegraphics[width=0.4\linewidth]{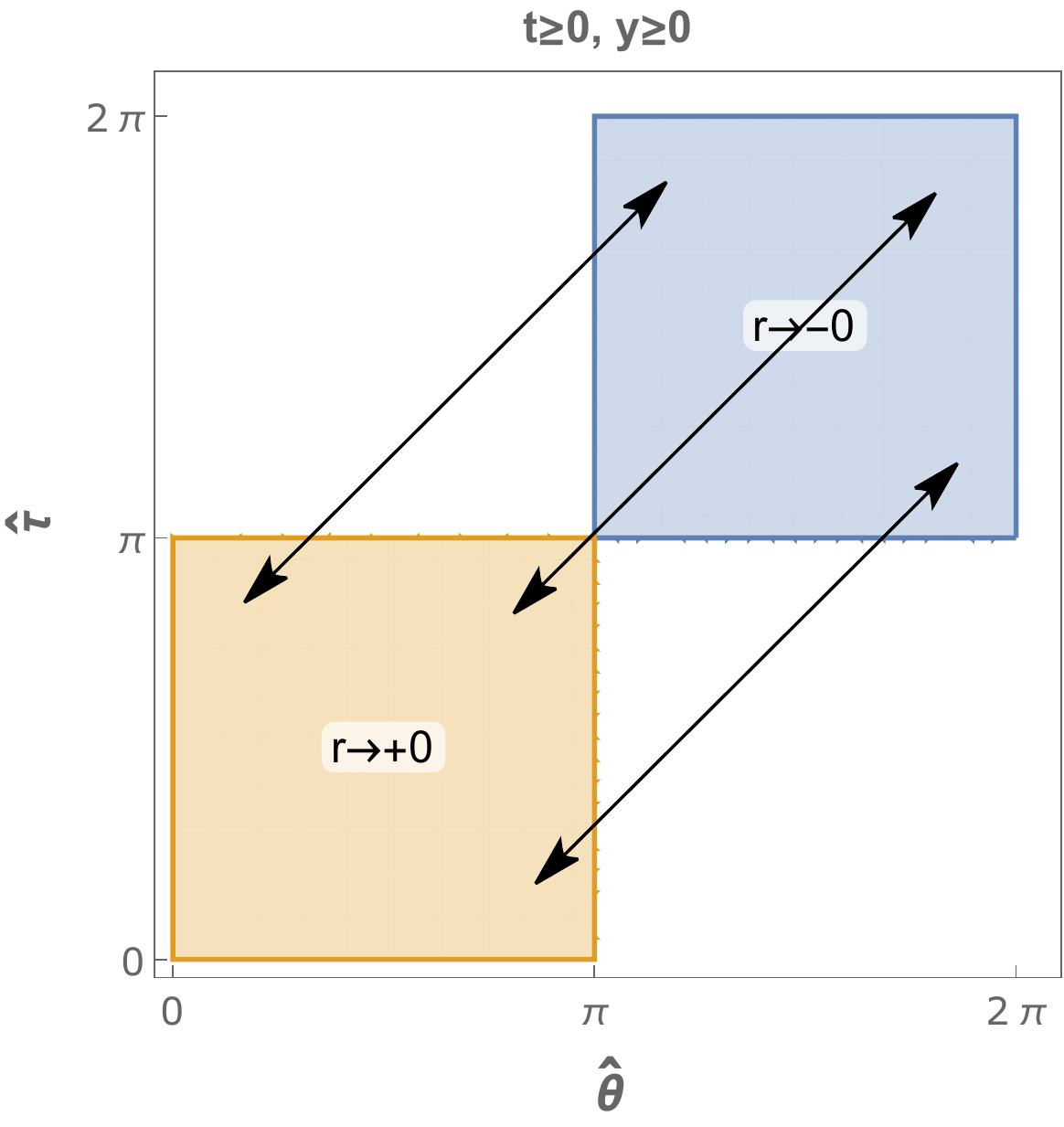}\quad
\includegraphics[width=0.4\linewidth]{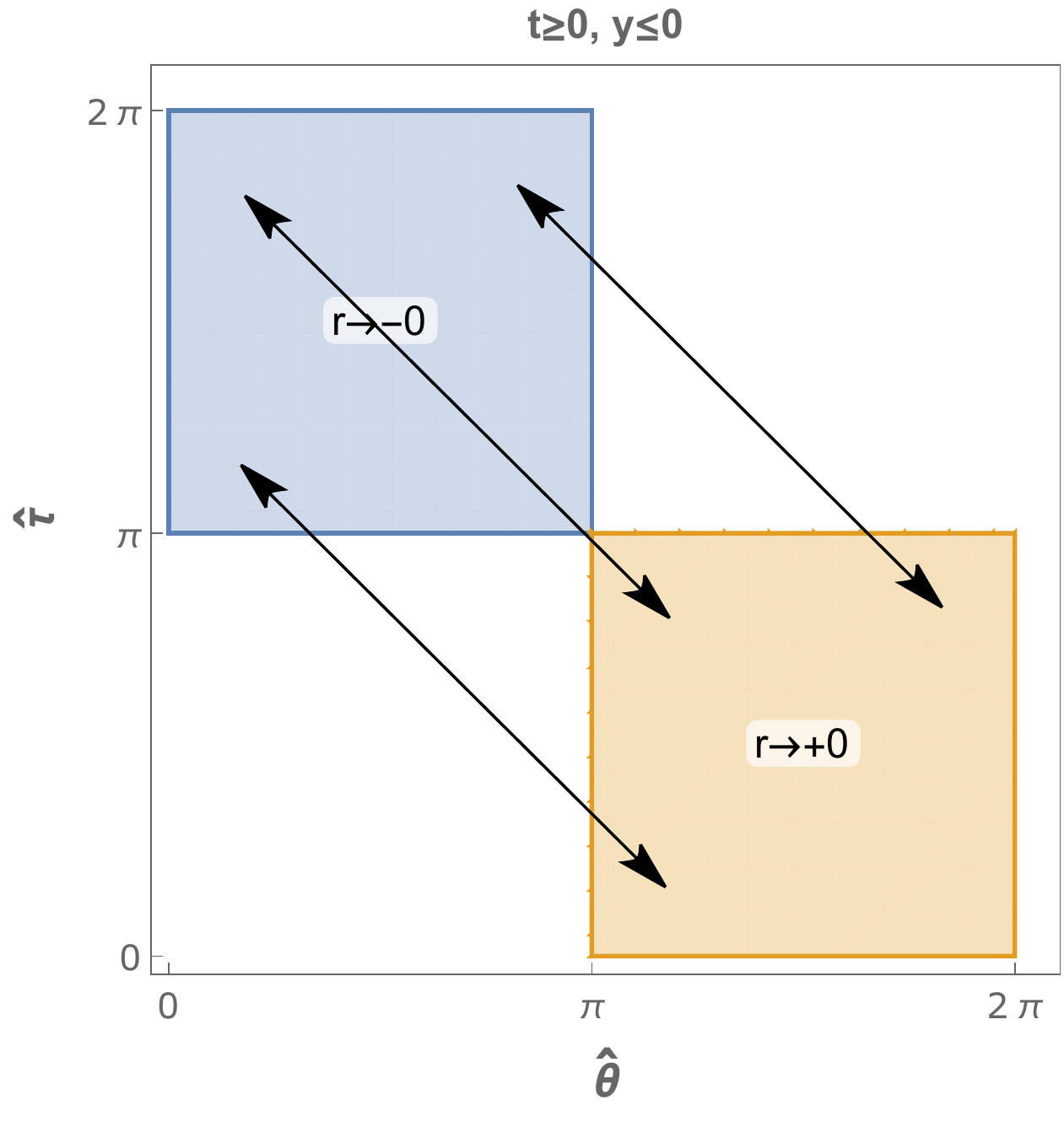}
\caption{Identification between points on the boundary of $\textrm{AdS}_3$}
\label{figure:bry_AdS}
\end{center}
\end{figure}

\begin{figure}[!h]
\begin{center}
\includegraphics[width=0.35\linewidth]{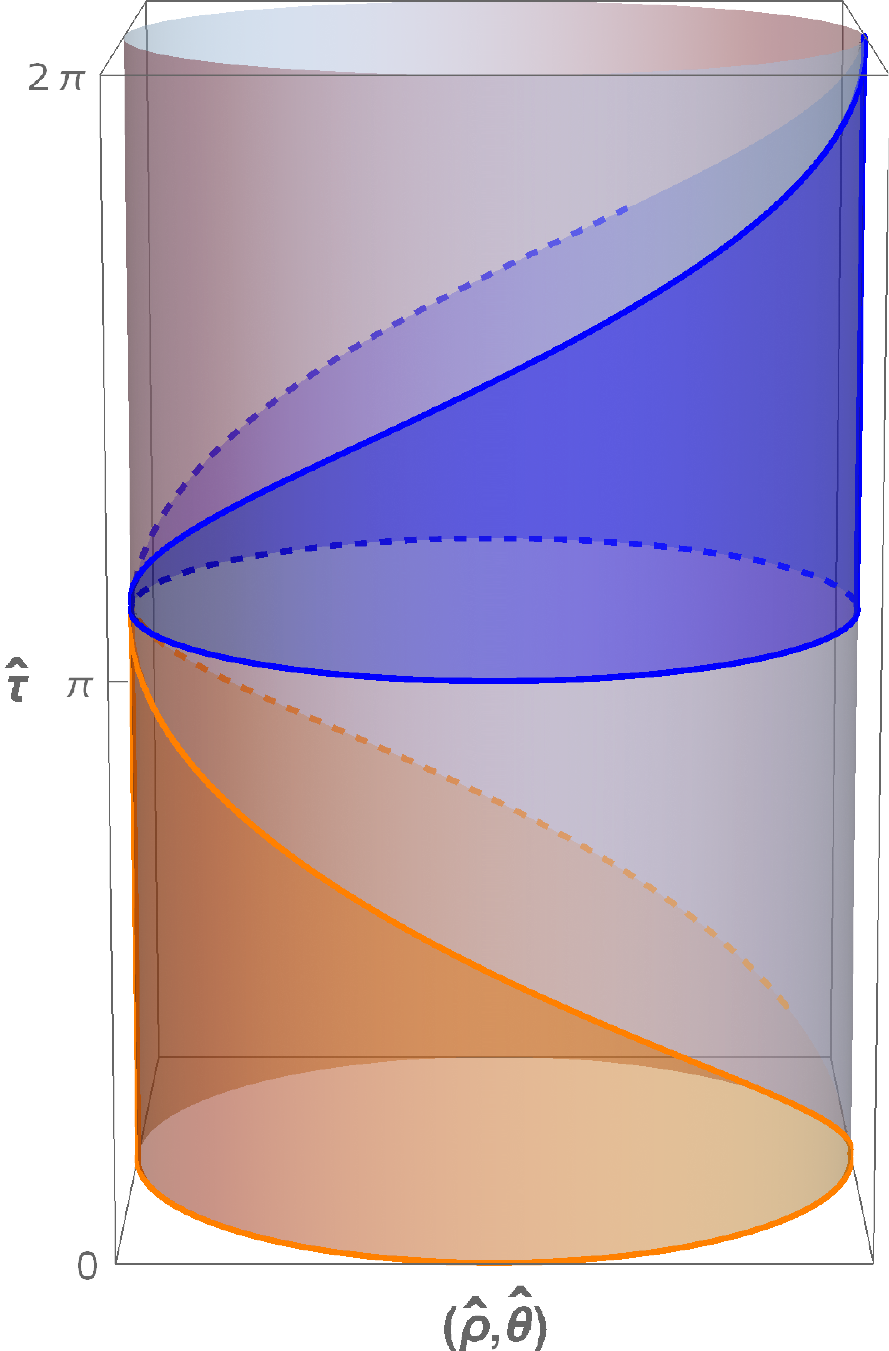}
\caption{Unconventional Poincare patch}
\label{figure:u_Poincare}
\end{center}
\end{figure}

\bibliographystyle{unsrt}
\bibliography{references}

\end{document}